\documentclass[aps,prl,showpacs,twocolumn,superscriptaddress]{revtex4-1}

\usepackage{latexsym}
\usepackage{color}
\usepackage{bm}
\usepackage{epsfig}
\usepackage{amsfonts}
\usepackage{amssymb}
\usepackage{amsmath}
\usepackage{graphicx}
\usepackage{euscript,tabularx}
\usepackage{textcomp}
\usepackage{longtable}
\usepackage{filecontents}

\begin{document}

\title{Time-resolved observation of fast domain-walls driven by vertical spin currents in short tracks}

\author{Joao Sampaio}
\author{Steven Lequeux}
\affiliation{
Unit\'e Mixte de Physique CNRS/Thales and Universit\'e Paris-Sud 11, 1 Ave.~A.~Fresnel, 91767 Palaiseau, France.}%

\author{Peter J.~Metaxas}
\affiliation{
Unit\'e Mixte de Physique CNRS/Thales and Universit\'e Paris-Sud 11, 1 Ave.~A.~Fresnel, 91767 Palaiseau, France.}
\affiliation{
School of Physics, M013,  University of Western Australia, 35 Stirling Highway, Crawley WA 6009, Australia.}

\author{Andre Chanthbouala}
\affiliation{
Unit\'e Mixte de Physique CNRS/Thales and Universit\'e Paris-Sud 11, 1 Ave.~A.~Fresnel, 91767 Palaiseau, France.}%

\author{Rie Matsumoto}
\author{Kay Yakushiji}
\author{Hitoshi Kubota}
\author{Akio Fukushima}
\author{Shinji Yuasa}
\affiliation{
National Institute of Advanced Industrial Science and Technology (AIST) 1-1-1 Umezono, Tsukuba, Ibaraki 305-8568, Japan.}%

\author{Kazumasa  Nishimura}
\affiliation{Process Development Center, Canon ANELVA Corporation, Kurigi 2-5-1, Asao, Kawasaki, Kanagawa 215-8550, Japan}
\author{Yoshinori  Nagamine}
\affiliation{Process Development Center, Canon ANELVA Corporation, Kurigi 2-5-1, Asao, Kawasaki, Kanagawa 215-8550, Japan}
\author{Hiroki  Maehara} 
\affiliation{Process Development Center, Canon ANELVA Corporation, Kurigi 2-5-1, Asao, Kawasaki, Kanagawa 215-8550, Japan}
\author{Koji~ Tsunekawa}
\affiliation{Process Development Center, Canon ANELVA Corporation, Kurigi 2-5-1, Asao, Kawasaki, Kanagawa 215-8550, Japan}

\author{Vincent Cros}
\author{Julie Grollier}
\affiliation{
Unit\'e Mixte de Physique CNRS/Thales and Universit\'e Paris-Sud 11, 1 Ave.~A.~Fresnel, 91767 Palaiseau, France.}%

\begin{abstract}
We present time-resolved measurements of the displacement of magnetic domain-walls (DWs) driven by vertical spin-polarized currents in track-shaped magnetic tunnel junctions. In these structures we observe very high DW velocities (600 m/s) at  current densities below $10^7 A/cm^2$. We show that the efficient spin-transfer torque combined with a short propagation distance allows to avoid the Walker breakdown process, and achieve deterministic, reversible and fast ($\approx$ 1 ns) DW-mediated switching of magnetic tunnel junction elements, which is of great interest to the implementation of fast DW-based spintronic devices.
\end{abstract}

\maketitle

Spin transfer torque offers a way to displace magnetic domain walls (DWs) and reverse magnetic elements using only electric currents \cite{berger84,Freitas85}. The typical systems studied have been the nanotrack of soft ferromagnetic material where an in-plane current is injected to displace one or more DWs \cite{klaui03,grollier03,parkin08}. The required high current density and slow DW velocity, however, remain significant obstacles preventing their application in practical devices \cite{parkin08}. 
In magnetic tunnel junctions (MTJ), we have recently demonstrated that currents perpendicular to the film can drive DWs orders-of-magnitude more efficiently than the in-plane injection geometry typically studied \cite{Khvalkovskiy,Chanthbouala,Metaxas}. Through the tunnelling magnetoresistance effect (TMR), this structure enables probing fast DW dynamics and also provides an efficient reading mechanism for future devices. We have also shown previously that DWs in this system can be depinned in deterministic times by moderate current densities ($\approx$ 10$^7$ A/cm$^2$) and that they can travel at velocities of the order of 500 m/s \cite{Metaxas}, well above the typical observed velocity observed in current in-plane studies \cite{Hayashi07}. In the present letter, we present time resolved, non-averaged (i.e. single shot) transport measurements of the DW displacement at high velocity. We then investigate how such high-velocities can take place in conditions where the Walker breakdown process is expected to severely limit the DW mean velocity \cite{Walker}.

\begin{figure}[h]
   \centering
	\includegraphics[width=0.5\textwidth]{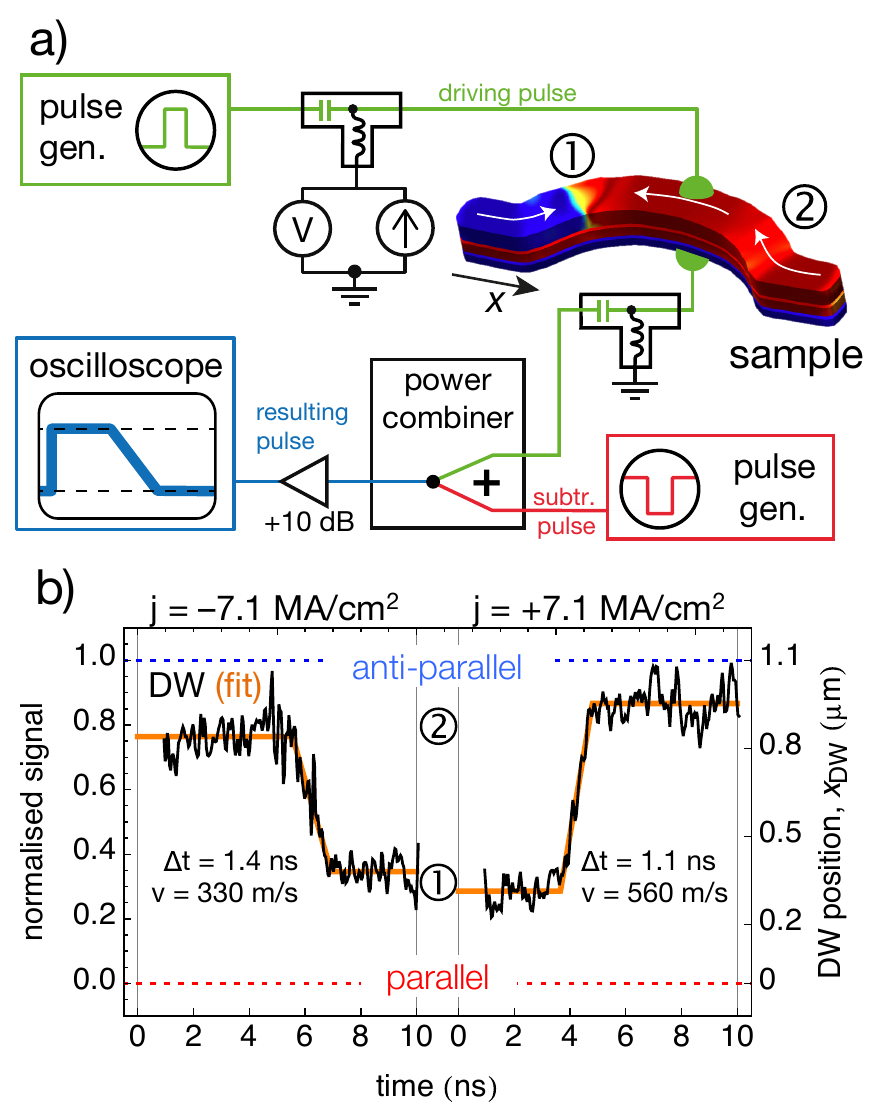}
     \caption{(a) Schematic of the time-resolved, non-averaged measurement setup, showing the curved track shape of the MTJ with the stable DW positions 1 and 2. (b) Time-resolved non-averaged measurements of a tail-to-tail DW displacement leftwards (left plot) and rightwards (right plot). The black line is the recorded signal normalized to the signals obtained in the fully parallel (red) and anti-parallel (blue) states. The right scale is the calculated DW position. The orange line is a linear fit, used to calculate the DW velocity and transition time. }
\label{fig1}
\end{figure}

The magnetic tunnel junction tracks were fabricated by electron beam lithography and ion beam milling from a film deposited by magnetron sputtering with structure PtMn 15/ CoFe 2.5/ Ru 0.9/ CoFeB 3/ MgO 1.17/ NiFe 5/ Ru 10 (thicknesses in nm) and a TMR ratio of 24\% (minimum resistance of $94 \Omega$). The CoFe/Ru/CoFeB tri-layer constitutes a synthetic anti-ferromagnet (SAF), and serves as both a reference layer for the TMR effect, and as a polarizer for the vertically injected current. The free layer, wherein the DW is created, is the Permalloy layer. As illustrated in Fig. \ref{fig1}(a), the 110 nm wide track is shaped as a wide arc so a DW can be injected in the free layer by applying a saturating field ($\approx$ 80 mT) transversally to the track (i.e. parallel to its width). The horizontal straight ends serve to geometrically pin the DW and impede it from leaving the track \cite{lewisCurve}. The position of the DW can be deduced from the sample resistance: as the DW changes position, the proportion of domains parallel and anti-parallel to the reference layer will change linearly. We deduce the DW position from a parallel resistor formula. We observed that the DW has two stable positions at both wire ends, labeled 1 and 2 in Fig. \ref{fig1}, which for a tail-to-tail DW are respectively at $x_{DW}$ = 0.24 and 0.82 $\mu$m. The DW can be switched back and forth between positions 1 and 2 by applying a small magnetic field, and also by applying short current pulses of a few MA/cm$^2$. 

To obtain a time-resolved measurement of the DW movement, the setup of Fig. \ref{fig1}(a) is used \cite{cui10}. The top MTJ electrode is connected to a pulse generator that produces the driving pulse (green circuit in Fig. \ref{fig1}(a)). A voltmeter and a current source are also connected for measuring the DC resistance before and after the pulse. The impedance mismatch between the driving circuit (Z = 50 $\Omega$) and the sample in series with the rest of the RF circuit (Z = R + 50 $\Omega$ $\approx$ 155 $\Omega$) causes the pulse to be partially reflected and transmitted. A second pulse of opposite polarity is then added to the transmitted pulse with help of a resistive power combiner (red circuit in Fig. \ref{fig1}(a)). The amplitude and shape of the second pulse is chosen so that the resulting pulse has an approximately zero average voltage, suitable to be subsequently amplified (+10 dB), and registered using an oscilloscope. In order to remove the effects of the incident pulse shape, the signals presented here are normalized with respect to two reference pulses obtained in the parallel and anti-parallel saturated states by application of a large field along the x axis:

\begin{equation}
s= \frac{s_{raw}-s_{P}}{s_{AP}-s_{P}}
\label{normSignal}
\end{equation}

where $s_{raw}$ is the registered signal, $s$ the normalized signal, and $s_{AP}$ and $s_{P}$ are the reference signals. The tail-to-tail DW position can then be approximated by $x_{DW} \approx l.s$, $l=1.1\mu m$ being the track length.

Due to the significant capacitance between the samples lithographically defined contacts, the signal cut-off frequency is $\approx$ 2.5-3 GHz, corresponding to $\tau$ $\approx$ 0.4 ns, which limits the minimum measurable transition time and causes an underestimation of the DW velocity. 

The left plot of Fig. \ref{fig1}(b) shows the normalized signal produced by a 10 ns, j = -7.1 MA/cm$^2$ pulse as a function of time after pulse arrival \cite{calcOfJ}. The DW position is indicated on the right axis. A small bias field (H$_{Bias}$= 0.6 mT) was applied to stabilize the two DW positions. Before the pulse, a tail-to-tail DW had been set at position 2 and after the pulse it was at the pinning site 1. A linear fit (orange line) was used to calculate the transition time $\Delta t$ and DW velocity $v$, which in this measurement is 1.4 ns and 330 m/s. Positive currents produced either no effect or a transition to the fully AP state (i.e. ejection of the DW). The reverse experiment is shown in the right plot of \ref{fig1}(b), with the DW initially at position 1 and a positive driving pulse of j = +7.1 MA/cm$^2$. It shows an opposite transition, with $\Delta t$ = 1.1ns and $v$ = 556 m/s. We attribute the difference in DW velocity with the direction of the current to small differences in the shape of the two pinning sites and to the asymmetric and inhomogeneous stray field produced by the fixed reference layer \cite{stray}.

\begin{figure}[h]
   \centering
	\includegraphics[width=0.5\textwidth]{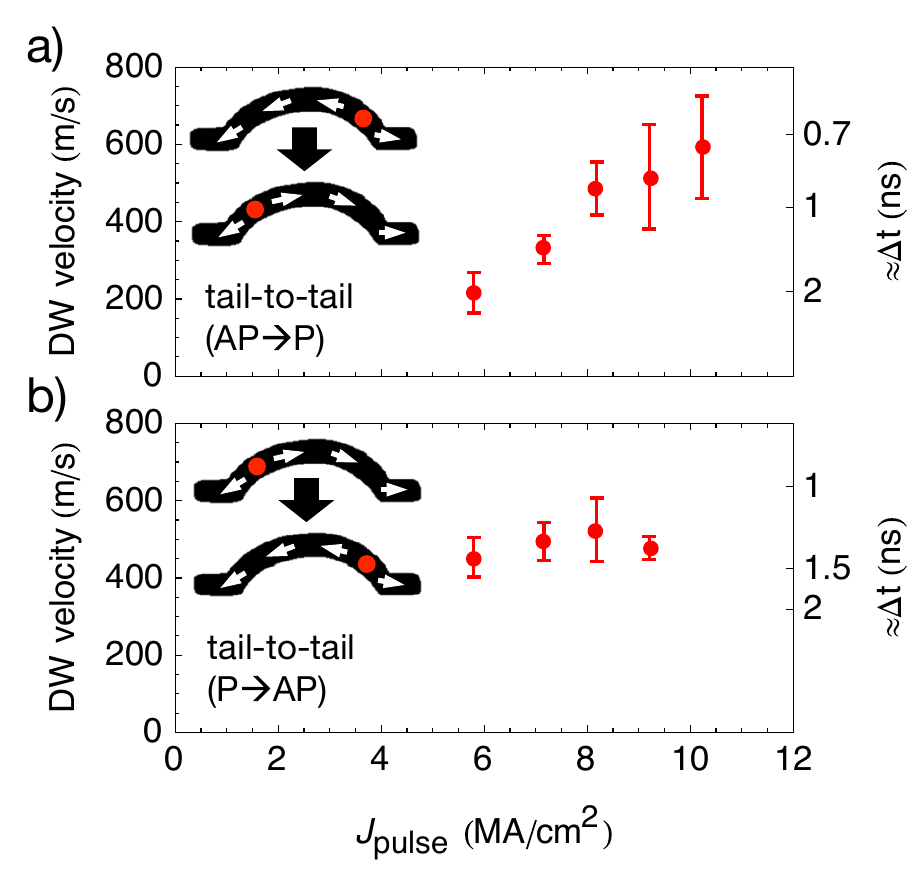}
     \caption{Velocity of the DW versus current density for the two DW polarities (HH and TT) and for two directions of propagation. The error bars represent the dispersion between measurements (n=4). The insets represent the initial and final magnetization states and DW positions.}
\label{fig2}
\end{figure}

The transition time $\Delta t$ and the DW velocity also change with current density. The average DW velocity as a function of current density is shown in Fig. \ref{fig2}(a), for a tail-to-tail DW initially in position 1. It can be seen that the velocity increases monotonically with the current density, within the error margin, up to velocities of the order of 600 m/s. No values are shown for lower current density, as they are insufficient to depin the DW. Fig. \ref{fig2}(b) shows the results for the tail-to-tail DW moving in the opposite direction. In this case, the DW velocity is about 500 m/s and shows no significant change with the current density. As seen before, the DW velocity for a given current density and the threshold current density for depinning vary with the sense of propagation. This is a consequence of the stray field from the reference layer and minute differences in the shape of the two pinning sites. Nevertheless, the measurements of \ref{fig2} for tail-to-tail DWs and measurements with a DW of the opposite polarity (head-to-head; not shown) reveal that there are aspects common to all measured cases. Firstly, we observe high DW velocities (between 400 and 600 m/s) at current densities below 10 MA/cm$^2$. This is a consequence of the high efficiency of the spin-transfer torque of vertically injected currents, as referred before \cite{Khvalkovskiy, Chanthbouala, Metaxas}. Secondly, we never observe a drop in DW mean velocity at higher driving currents, i.e. there is no sign of the Walker breakdown regime. The hallmark of the Walker regime is a precessional DW movement with periodic reversal of the DW propagation direction, and the associated dramatic drop in the average DW velocity \cite{Walker,lee07,glathe08}. Different mechanisms to suppress the Walker regime have been proposed and demonstrated \cite{lewisComb,glathe08,nakatani03}, resulting in higher DW velocities. The Walker regime is observed when the driving torque exceeds a threshold (the Walker field or equivalent current). The absence of Walker breakdown is equally confirmed in the individual measurements (as in Fig. \ref{fig1}(b)), where no velocity reversals can be discerned.

To understand the nature of these fast DW transitions and the absence of the Walker precession, we perform a micromagnetic study of DW propagation in short lengths driven by spin-tranfer torque. We start by studying a straight track, and then compare it to simulations of the actual structure shape. We use the OOMMF code \cite{oommf} with the following parameters: cell-size 5x5x5 nm$^3$, $M_s$ = 0.8 MA/m, A = 13 pJ/m, current polarization of 0.4 and out-of-plane to in-plane torque ratio of 0.3. We observe that only the out-of-plane torque has a significant action on the DW \cite{Khvalkovskiy, Chanthbouala}.

\begin{figure}[h]
  \centering
	\includegraphics[width=0.5\textwidth]{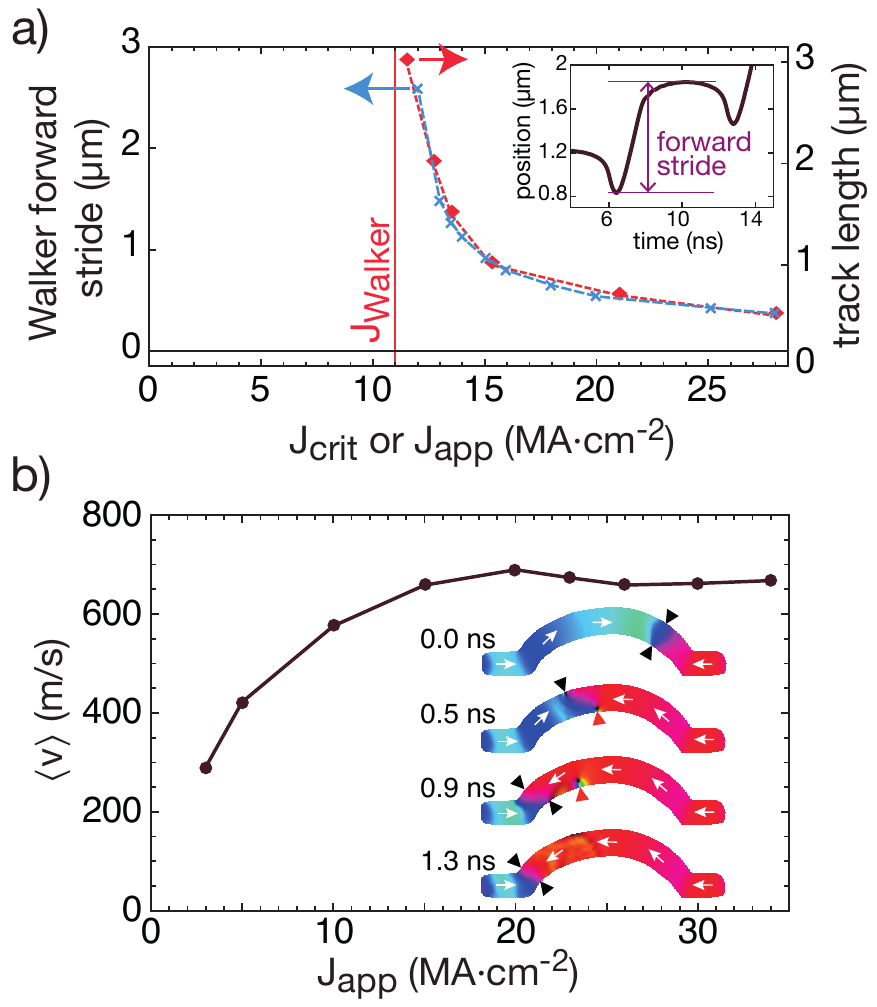}
  \caption{(a) Comparison of the Walker forward stride versus applied current density J$_\text{app}$ for the steady state propagation (blue crosses), and the track length versus critical current density J$_\text{crit}$ for the propagation in a limited track (orange squares). The vertical scales are shifted to draw attention to the similarity of the two curves. Inset: position versus time of a DW in the Walker regime, illustrating the Walker forward stride.
(b) Average velocity vs applied current of a DW propagating in a track of the same shape as the experimental device. Inset: magnetization snapshots for J$_\text{app}$ = 34 MA.cm$^{-2}$ at different times. The black triangles indicate the DW position at the track borders, and the red triangles the injected anti-vortex. The colours represent the in-plane magnetization angle (cyan for +x, red for -x, purple for +y and green for -y). }
\label{fig3}
\end{figure}

Firstly we study the steady propagation state of current-driven DW in a permalloy stripe of 5 x 100 nm$^2$ cross-section (similar to our sample) and 3 $\mu$m long, initialized with a DW 0.2 $\mu$m from the border. After an initial short transient regime, the DW either propagates at constant velocity (for an applied current density $J_\text{app} < J_\text{Walker} \approx$ 11 MA.cm$^{-2}$) or with periodical oscillations of the instant velocity (for $J_\text{app} > J_\text{Walker}$), as illustrated in the inset of \ref{fig3}a. The forward stride of the DW motion during one Walker oscillation, i.e. how much a DW moves forward before reversing its velocity, is shown as a function of applied current in the blue line and crosses of Fig. \ref{fig3}(a). As expected, it decreases with increasing applied current. 

We will now consider the propagation of a DW in a track of limited length, and study the velocity of the current-driven DW to the opposite end of the track. In all cases, we observe a critical current density $J_\text{crit}$ above which the mean velocity drops from $\approx$ 600 to $\approx$ 100 m/s. The evolution of $J_\text{crit}$ with track length is shown in the orange line and squares of Fig. \ref{fig3}(a). For longer tracks, $J_\text{crit}$ asymptotically approaches $J_\text{Walker}$. For shorter tracks, $J_\text{crit}$ is significantly larger, and is about 28 MA.cm$^{-2}$ for 0.5 $\mu$m long track, i.e. 2.5 times larger than $J_\text{Walker}$. This can be interpreted by comparing the track length to the Walker forward stride (orange and blue lines of Fig. \ref{fig3}(a), respectively.) This plot shows that the high mean velocity in the limited tracks (below $J_\text{crit}$) occurs when the track is shorter than the Walker forward stride, causing the DW to reach the end of the track before it starts its retrograde motion. 

Finally, we simulate the propagation of DW in a track shaped as our structure (Fig. \ref{fig1}(a)). The stray field from the SAF reference layer was simulated separately and included as a static field. As is the case for the experimental structure, we observe two stable DW positions on either side of the arc section (see insets of \ref{fig3}(b), and we considered the transitions between these positions. The mean velocity versus driving current density is shown in Fig. \ref{fig3}(b). We observe no motion below 3 MA.cm$^{-2}$ due to the pinning. Above it, the velocity increases up to a threshold current density (~15 MA.cm$^{-2}$) where it saturates at 650 m/s. In all simulations above 20 MA.cm$^{-2}$, an anti-vortex core is nucleated in the DW, which is characteristic of the onset of the Walker breakdown process. The insets of Fig. \ref{fig3}(b) show snapshots of the magnetization during this process, with the anti-vortex indicated by a red triangle. This anti-vortex can detach itself from the DW and it eventually disappears through the track border. The curvedness of the track prevents a direct quantitative comparison to the straight track model presented above. However, the same features can be identified. For low fields we observe the sub-Walker regime with increasing velocities. Above a threshold current density, the Walker process is initiated (a vortex or an anti-vortex core is injected), but the DW reaches the final pinning position before a reversal of the propagation direction occurs, resulting in an elevated mean velocity. This threshold current density (15 MA.cm$^{-2}$) is comparable to the critical current density of the straight limited track for the same propagation distance (21 MA.cm$^{-2}$ for a 0.7 $\mu$m long track). 

We have shown that vertically-injected currents can reverse an MTJ track using DWs displacement tracks in a very short time ($<$ 1 ns) and at moderate current density ($\approx$ 10$^7$ A/cm$^2$). This is possible due to the suppression of the Walker precession and the subsequent orders-of-magnitude increase in mean DW velocity, an effect which is directly connected to a limited DW propagation distance. This effect causes the reversal time of a shorter track segment to be doubly decreased: less distance to travel at higher mean velocity. 

These systems can serve as efficient memory elements. Comparing to spin transfer torque magnetic random access memories (STT-MRAMs) with uniform magnetization \cite{khvSTTMRAM}, the described system has the advantage of allowing multi-bit storage by introducing multiple stable DW positions. The presence of a DW, however, limits the size of the structure to at least a few DW lengths, a limitation that can be minimized by using perpendicularly magnetized materials with thinner DWs. Additionally, the deterministic switching at moderate current densities allows faster operation speeds than in conventional STT-MRAM, where stochastic switching times can limit the device speed and introduce writing errors \cite{khvSTTMRAM}. Concluding, short magnetic nano-tracks reversed by current-driven DWs are promising memory elements for future fast spintronic devices. 
 
\subsection*{Acknowledgements}
The authors acknowledge financial support from the European Research Council (Starting Independent Researcher Grant No.~ERC 2010 Stg 259068), the French ANR grant ESPERADO 11-BS10-008 and the French Ministry of Defense (DGA).


\begin{thebibliography}{1}
\bibitem{calcOfJ}
The current density was calculated from transmitted signal amplitude: $I_\text{sample} = \frac{V_\text{transmitted}}{Z_0} $, where $Z_0 = 50 \Omega $ is the impedance of the cable.
\end{thebibliography}
\end{document}